\documentclass[showpacs,preprint]{revtex4}

\usepackage{graphicx}
\usepackage{dcolumn}
\usepackage{amsmath}

\makeatletter
\def\btt#1{\texttt{\@backslashchar#1}}
\DeclareRobustCommand\bblash{\btt{\@backslashchar}} \makeatother

\begin{document}


\title{Shadow of Schwarzschild-Tangherlini black holes}
\author{Balendra Pratap Singh}
\email{balendra29@gmail.com}
\affiliation{Centre for Theoretical Physics,
 Jamia Millia Islamia,  New Delhi 110025
 India}

 \author{Sushant G. Ghosh}
\email{sghosh2@jmi.ac.in}
\affiliation{Centre for Theoretical Physics  and Multidisciplinary Centre for Advanced Research and Studies (MCARS),
 Jamia Millia Islamia,  New Delhi 110025
 India}
\affiliation{Astrophysics and Cosmology Research Unit,
 School of Mathematics, Statistics and Computer Science,
 University of KwaZulu-Natal, Private Bag X54001,
 Durban 4000, South Africa}

\begin{abstract}
We study the shadow cast by the H$D$ Schwarzschild-Tangherlini  black hole, and analytically calculate the influence of extra dimensions on the shadow of a black hole. 	A black hole casts a shadow as an optical appearance because of its strong gravitational field which is known to be a dark zone covered
by a  circle for a Schwarzschild black hole.   We demonstrate that the null geodesic equation can be integrated by Hamilton-Jacobi approach, which enable us  to investigate the shadow cast by the H$D$ Schwarzschild-Tangherlini  black holes.   Interestingly, it turns out that, for fixed values of the mass parameter, the shadow in H$D$ spacetimes are  smaller  when compared with 4$D$ Schwarzschild  black hole. Further, the shadows of H$D$ Schwarzschild-Tangherlini  black holes are concentric circles with radius of the circle decreases with increase in $D$. We visualize the photon regions and the shadows in various dimensions for different  values of the parameters, and the energy emission rates are is also investigated.  Our results, in the limit $D=4$, reduced exactly	to  \emph{vis-$\grave{a}$-vis} Schwarzschild black hole case.
\end{abstract}
\maketitle

\section{Introduction}
There is a great interest to investigate nature of the black holes, i.e., mass and spin of the black hole,  which  can be possibly determined by observation of black hole shadow  \cite{Vries,Takahashi, Bambi, Kraniotis}. Now, it is a general belief that a black hole, if it is in front of a luminous background, will cast a shadow. To observe the shadow of the black hole at the center of the Milky way, one will be looking for a  ring of light around a region of darkness, which is called the black hole’s “shadow.” That light is produced by matter that is circling at the very edge of the event horizon, and its shape and size are determined by the black hole’s mass and spin.  A shadow is the optical appearance cast by a black hole, and its existence was first studied by Bardeen \cite{JMB}. The shadow of a nonrotating black hole is circular \cite{Bozza:2009yw} while a distorted circle for rotating black holes and this is due to the presence of the spin parameter  \cite{JMB,Chandrasekhar:1992,Falcke:1999pj}. It was Synge \cite{Synge:1966} who studied shadow of  Schwarzschild black holes and thereafter  Luminet \cite{Luminet:1979} discussed the optical properties of static, spherically symmetric  Schwarzschild black holes and constructed a simulated photograph of the shadow. Later, it received a significant attention and has become a quite active research field (for a review, See \cite{bozzareview}). More discussion on shadow of  Schwarzschild black hole \cite{Distort}, other spherically symmetric black holes \cite{Perlick:2015vta} have been intensively studied and have been extended to  rotating black holes \cite{Hioki:2009na, Takahashi:2005hy, Wei:2013kza, Abdujabbarov:2012bn, Amarilla:2011fx,Amarilla:2013sj,Atamurotov:2013sca, Grenzebach:2014fha, Amir:2016cen, Abdujabbarov:2015xqa, Atamurotov:2015nra, Abdujabbarov:2016hnw}. 

Over the past decade there has been an increasing interest in the
study of black holes, and related objects, in H$D$, motivated to a large extent by developments in string
theory. Recently, it has been proposed that our universe may have emerged from a black hole in a higher-dimensional (H$D$) Universe.  Black holes are very interesting gravitational as well as geometrical objects to study in 4$D$ which may also exist in H$D$ spacetimes.  Although,  at present, the  work on H$D$ can probably be most fairly described as extended theoretical speculation and it has no direct observational and experimental support, in contrast to 4$D$ general relativity. However, this theoretical work has led to the possibility of proving the existence of extra dimensions which is best demonstrated by  Reall and Emparan to show that there is a 'black ring' solution in 5$D$ \cite{Emparan}. If such a 'black ring' could be produced in a particle accelerator such as the Large Hadron Collider, this would provide the evidence that H$D$ exist.   There are  other reasons for this interest in  H$D$ black holes \cite{Ghosh:2008zza, Ghosh:2001pv,Ghosh,Ghosh:2001fb,Ghosh:2003ku} in particular, e.g., the statistical calculation of black hole entropy using string theory was first done for certain 5$D$ black holes \cite{Strominger:1996sh} and also  the possibility of producing tiny H$D$ black holes at LHC in certain ”brane-world” scenarios \cite{Kanti}. This study of shadow is extended to H$D$ spacetime by several researchers, e.g., Papnoi et. al.  \cite{Papnoi:2014aaa} have studied the shadow cast by 5$D$ rotating Myers-Perry black holes, for pure Gauss-Bonnet  gravity rotating black holes by \cite{Abdujabbarov:2015rqa}, also in Kaluza-Klein gravity \cite{Amarilla:2013sj}.   However, the results of these work can't go over to the Schwarzschild black hole. 

Hence, it is pertinent to investigate the apparent shape of  H$D$ Schwarzschild-Tangherlini black holes to visualize the shape of the shadow and compare the results with images for  4$D$  Schwarzschild black hole.  An apparent Shape of black hole is determined via boundary of the shadow which can be studied by  the null geodesic equations. Clearly, the extra spacetime dimension shall change the equations of motion  which may lead to the modification of black hole shadow.  

The paper is organized as
follows: in Sect.~\ref{sect2}, we review the Schwarzschild-Tangherlini
black hole solutions and present the associated thermodynamical quantities.  In  Sect.~\ref{sect3}, we have presented the particle
motion around the Schwarzschild Tangherlini black hole by using the Hamilton-Jacobi approach necessary to discuss the shadow.  The observables are introduced Sect.~\ref{sect4} to plot the apparent shapes of the black hole shadows  and finally in Sect.~\ref{sect6}, we have concluded with the results.

\section{The Schwarzschild-Tangherlini spacetime \label{sect2}}
We present the basic framework for general relativity in H$D$, and introduce the Schwarzschild Tangherlini solutions that generalize the 4$D$ Schwarzschild solution to the H$D$. The action in H$D$ space-time
\begin{equation}
 \mathcal{I}=\int d^Dx\sqrt{-g}R .
\label{emd}
\end{equation}
This is a straightforward generalization of Einstein-Hilbert action to H$D$ and the only aspect that deserves some attention is the implicit definition of Newton's constant $G_D$ in H$D$, without loss of generality we use units such that $G_D=c=1$. Using variational principle, one obtains the Einstein equation in H$D$ as 
\begin{equation}\label{feqn}
R_{ab}-\frac{1}{2}Rg_{ab}=0 \quad \mbox{or} \quad  R_{ab}=0,
\end{equation} 
where $R_{ab}$, $R$ and $g_{ab}$ are respectively Ricci  tensor, Ricci scalar  and metric tensor. Tangherlini {\cite{st}} has found the asymptotically flat, static and spherically symmetric vacuum solution of (\ref{feqn}) as a generalization of the
Schwarzschild black hole 
\begin{equation}\label{metric}
	ds^2=-\left(1-{\frac{\mu}{r^{D-3}}}\right) dt^2+\frac{1}{\left(1-{\frac{\mu}{r^{D-3}}} \right) }dr^2+r^2d\Omega^2_{D-2},
\end{equation}
where
\begin{equation}
  d\Omega^2_{D-2}=d{\theta_1}^2+{\sin^2{\theta_1}^2}{d{\theta_2}^2}+, . . . ,+\prod_{i=1}^{D-3}\sin^2\theta_id{\theta}_{D-2}^2,
\end{equation}
is a metric on $(D-2)$-dimensional unit sphere. The parameter $\mu$ is related to black hole mass $M$ via
 \begin{equation}
\mu=\frac{16{\pi}M}{(D-2)\Omega_{D-2}},
\end{equation}
where $\Omega_{D-2}$ is the volume of the $(D-2)$ dimensional sphere given by
\begin{equation}
	\Omega_{D-2}=\frac{2\pi^{\frac{D-1}{2}}}{\Gamma\left(\frac{D-1}{2}\right)}.
\end{equation}
This suggests that the Schwarzschild solution generalizes to H$D$ in the form given by the metric (\ref{metric}). First thing to be noticed is that this simplifies to the normal Schwarzschild metric
when $D=4$. Secondly, the H$D$ version is very similar to the
4$D$ one
and the notable difference is when we go to H$D$ then the fall off term $1/r$ replaced with the $1/{r^{D-3}}$.  As shown by Tangherlini \cite{st}, this turns out to give the correct solution and that the metric (\ref{metric}) is indeed Ricci flat, and  this solution is  called the Schwarzschild-Tangherlini
solution.   As can be seen from equation  (\ref{metric})  we have no black-hole solution in
$ D= 3 $. If the mass parameter $\mu < 0$ then we get a naked singularity, which is not
physical. If $\mu > 0$,
the black hole horizon radius $r_h$  which is obtained by solving $g_{tt}(r_h)=0$ as 
\begin{equation}
	r_h=\left[\frac{16{\pi}M}{(D-2)\Omega_{D-2}}\right]^{\frac{1}{D-3}}.
\end{equation} 
The event horizon is the nonrotating Killing
horizon and its spatial cross-section is around $(D-2)$-sphere.
The mass of the black hole in terms of the horizon radius $r_h$ gives
\begin{equation}
M=\frac{(D-2)\Omega_{D-2}}{16{\pi}}{r_h}^{D-3}.
\end{equation}
The area of the event horizon for the metric (\ref{metric}), is
\begin{equation}
A=\Omega_{D-2}r_h^{D-2}.
\end{equation}
The black hole entropy is expected to obey area law \cite{Frolov}. Explicitly the entropy $S$ in terms of $M$, can be written as
\begin{equation}
S={\frac{\Omega_{D-2}}{4}}\left[\frac{16 \pi  M}{(D-2)\Omega_{D-2}}\right]^{\frac{D-2}{D-3}}.
\end{equation}
The black hole obeys the first law of thermodynamics $dM=TdS$, which can be use to obtain Hawking Temperature as
\begin{equation}
T_{BH}=\frac{D-3}{4\pi r_h}.\label{T}
\end{equation} 
Clearly, all above result reduces to the 4$D$ Schwarzschild black holes when $D=4$ \cite{Frolov}. 
Although a black hole is invisible, it can cast a shadow when it is in front of a bright object. The aim of this work is to discuss the effect of extra dimension in a black hole shadow in the background of Schwarzschild-Tangherlini.
\section{Motion of a test particle \label{sect3}}
When a black hole is in front of the light source, the light reaches the observer when deflect due to gravitational field of the black hole. However, it turns out that some of the photons may fall into the black hole, which result a dark zone called the shadow, and the apparent shape of the black hole is the boundary of the shadow. We present the necessary calculations for obtaining the shape of Schwarzschild-Tangherlini black hole shadow, which demands the study of the motion of test particle. We employ the Lagrangian and Hamilton-Jacobi equation to obtain equations of motion, which requires a study of geodesics equation of a particle near Schwarzschild-Tangherlini black hole. We begin with the Lagrangian which reads
\begin{equation}
		\mathcal{L} = \frac{1}{2} g_{\mu \nu}\dot{x}^{\mu}\dot{x}^{\nu},
\end{equation}
where an over dot is derivative with respect to affine parameter $\tau$ and $g_{\mu\nu}$ is the metric tensor.
The canonically conjugate momentum for the Schwarzschild-Tangherlini black holes metric (\ref{metric}) can be calculated as
\begin{eqnarray}
&& P_{t}=\left(1-{\frac{\mu}{r^{D-3}}}\right)\dot{t}=E, \label{p1}\\
&&{P_r}= \left(1-{\frac{\mu}{r^{D-3}}}\right)^{-1}\dot{r}, \label{p3}\\
&&P_{\theta_i}=r^2\sum_{i=1}^{D-3}\prod_{n=1}^{i-1}\sin^2\theta_n\dot
{\theta}_{D-3}, \quad \:\:\:  i=1, . . . ,D-3 \label{p4}\\
&& P_{\phi}=r^2\prod_{i=1}^{D-3}\sin^2\theta_i\dot{\theta}_{D-2}=L, \label{p2}
\end{eqnarray}
where $P_\phi=P_{\theta_{D-2}}$  \cite{Vasudevan:2004ca}, $E$ and $L$ are  respectively energy  and angular momentum of the test particle. For $D=4$, we have $i=1$ and the quantities $P_{\theta_{i}}$ and $P_{\phi}$  reduces to Schwarzschild case, can be reads 
\begin{eqnarray}
&&P_{\theta_{1}}=r^2\dot{\theta_{1}},\nonumber\\ 
&&P_{\phi}=r^2\sin^2\theta_{1}\dot{\theta_{2}}.\nonumber
\end{eqnarray}
We use Hamilton-Jacobi method to  analyze photon orbits around the black hole and use the formulation of geodesic equations by Carter approach  for Schwarzschild black hole \cite{carter}, which we extended here to H$D$. The Hamilton-Jacobi method in the H$D$ reads 
\begin{equation}
\frac{\partial S}{\partial {\tau}}  =\mathcal{H}=-\frac{1}{2}  g^{\mu\nu}\frac{\partial S}{\partial 		x^\mu}\frac{\partial S}{\partial x^\nu}, \label{j}
\end{equation} 
where $S$ is the Jacobi action. On using (\ref{metric}) in Eq.~(\ref{j}), we obtain
\begin{eqnarray}
-2\frac{\partial S}{\partial {\tau}} &=& -\frac{1}{\left(1-{\frac{\mu}{r^{D-3}}}\right)}\left(\frac{\partial S_t}{\partial t}\right)^2+ \left(1-{\frac{\mu}{r^{D-3}}}\right)\left( \frac{\partial S_r}{\partial r}\right)^2 + 
\sum_{i=1}^{D-3}\frac{1}{r^2\prod\limits_{n=1}^{i-1}\sin^2\theta_n}\left(\frac{\partial S_{\theta_i}}{\partial {\theta}_i }\right)^2
\nonumber \\
&+& \frac{1}{r^2\prod\limits_{i=1}^{D-3}\sin^2\theta_i}\left(\frac{\partial S_{\phi}}{\partial {\phi}}\right)^2.
\end{eqnarray}
Here, we consider a additive separable solution for Jacobi action $S$, which can be expressed as
\begin{equation}
S= \frac{1}{2}{m^2} \tau -Et+ L\phi +
S_r (r) + \sum\limits_{i=1}^{D-3} S_{\theta_i}(\theta_i),\label{s}
\end{equation}
where $S_r(r)$ and $S_{\theta_i}(\theta_i)$ are respectively functions of $r$ and $\theta_i$ and $m$ is the mass of the test particle, which is zero for the photon. 
On the right hand side, the second and third term related with the conservation of energy and the angular momentum respectively.
 The Hamilton-Jacobi Eq.~(\ref{j}), on using (\ref{s}) can be recast as
\begin{eqnarray}
&&r^4\left(1-{\frac{\mu}{r^{D-3}}}\right)^2\left(\frac{\partial S_r}{\partial r}\right)^2=E^2r^4-r^2\left(1-{\frac{\mu}{r^{D-3}}}\right)\left(\mathcal{K}+{L}^2\right),\label{sr} \\
&&{\displaystyle\sum_{i=1}^{D-3}\frac{1}{\displaystyle\prod_{n=1}^{i-1}\sin^2\theta_i}}\left(\frac{\partial S_{\theta_i}}{\partial \theta_i}\right)^2=\mathcal{K}-\prod_{i=1}^{D-3}{L}^2\cot^2\theta_i,\label{stheta}
\end{eqnarray}
\begin{figure*}
    \begin{tabular}{c c c c}
	\includegraphics[scale=0.6]{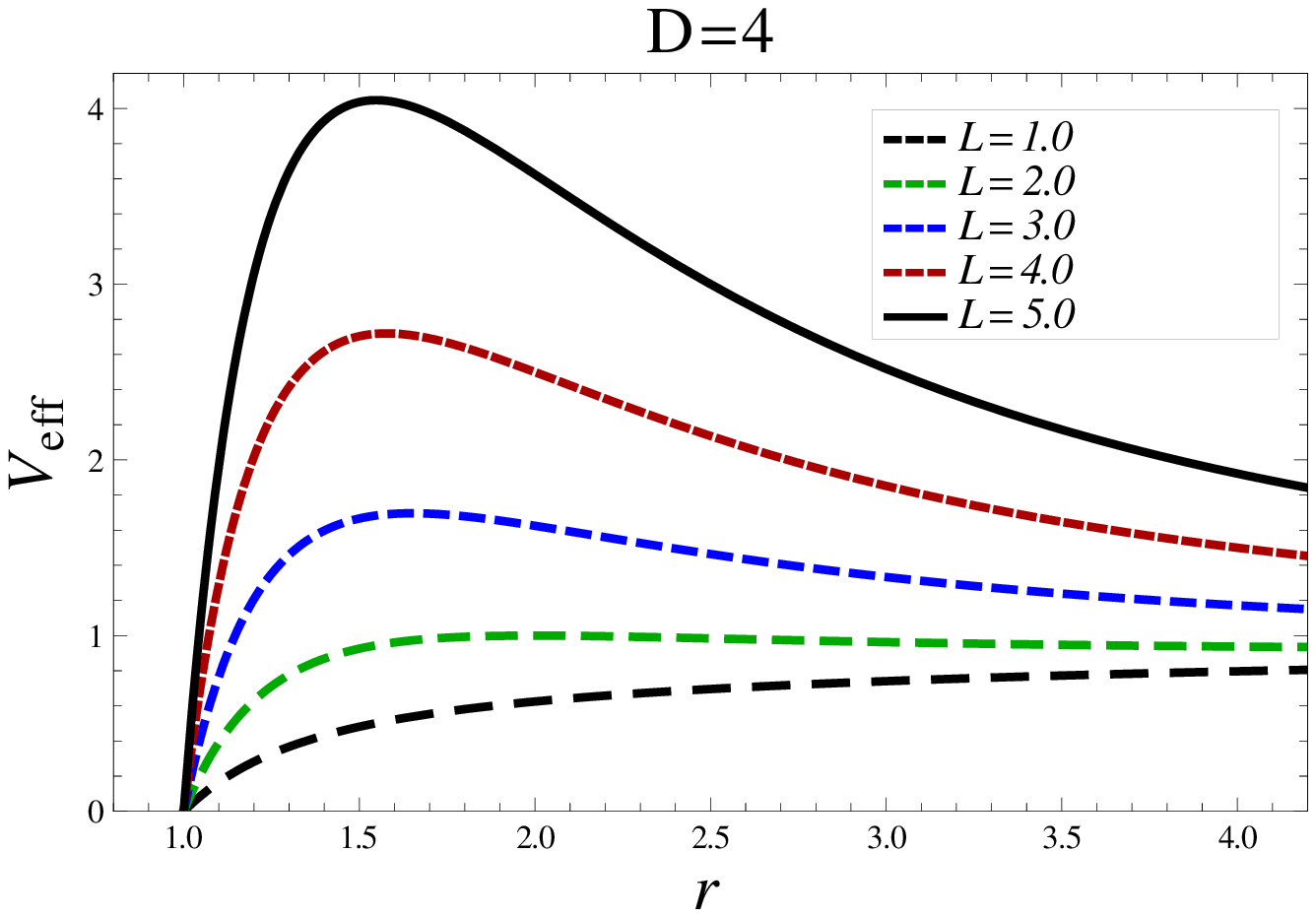} 
	\includegraphics[scale=0.6]{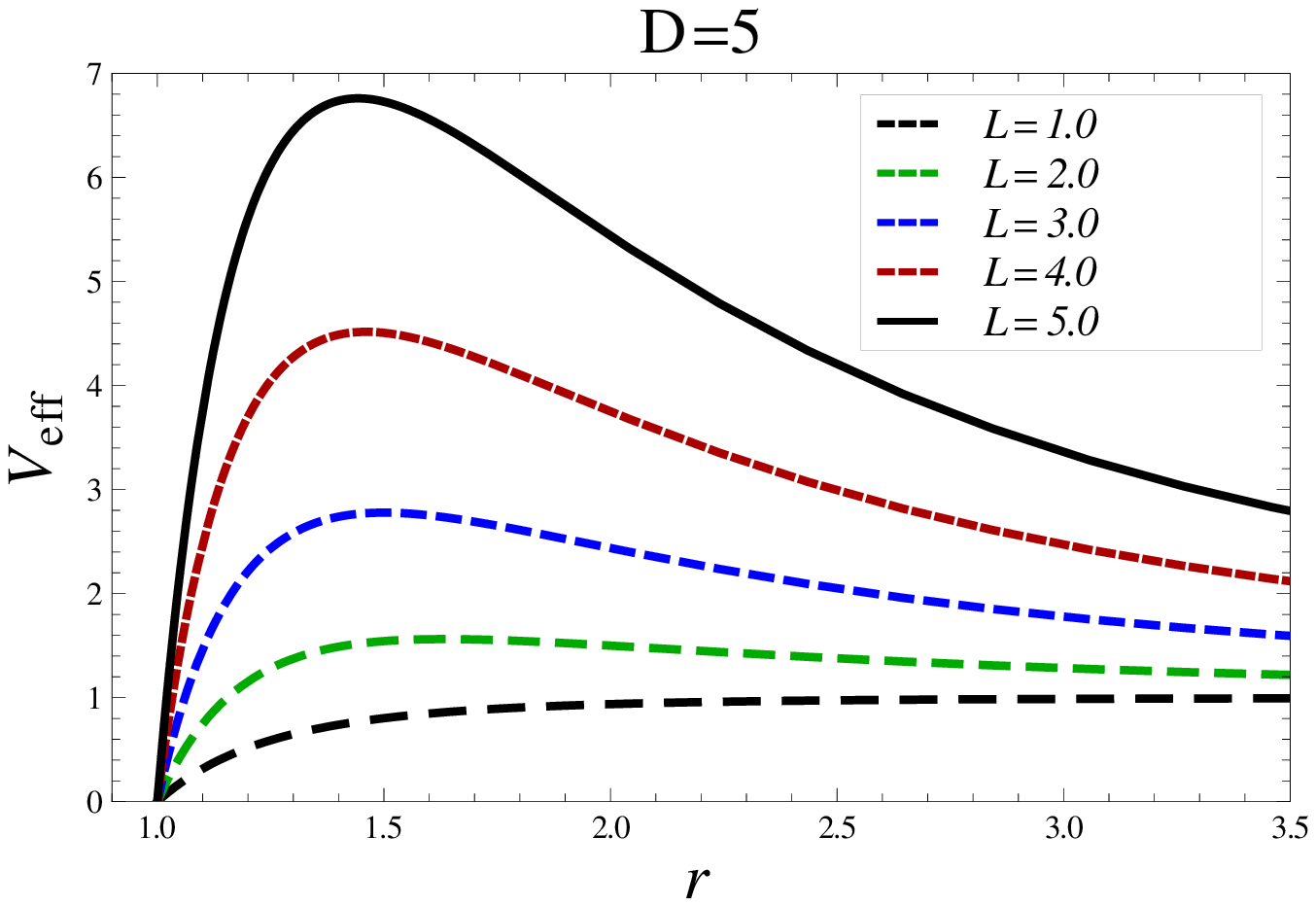} \\
     \includegraphics[scale=0.6]{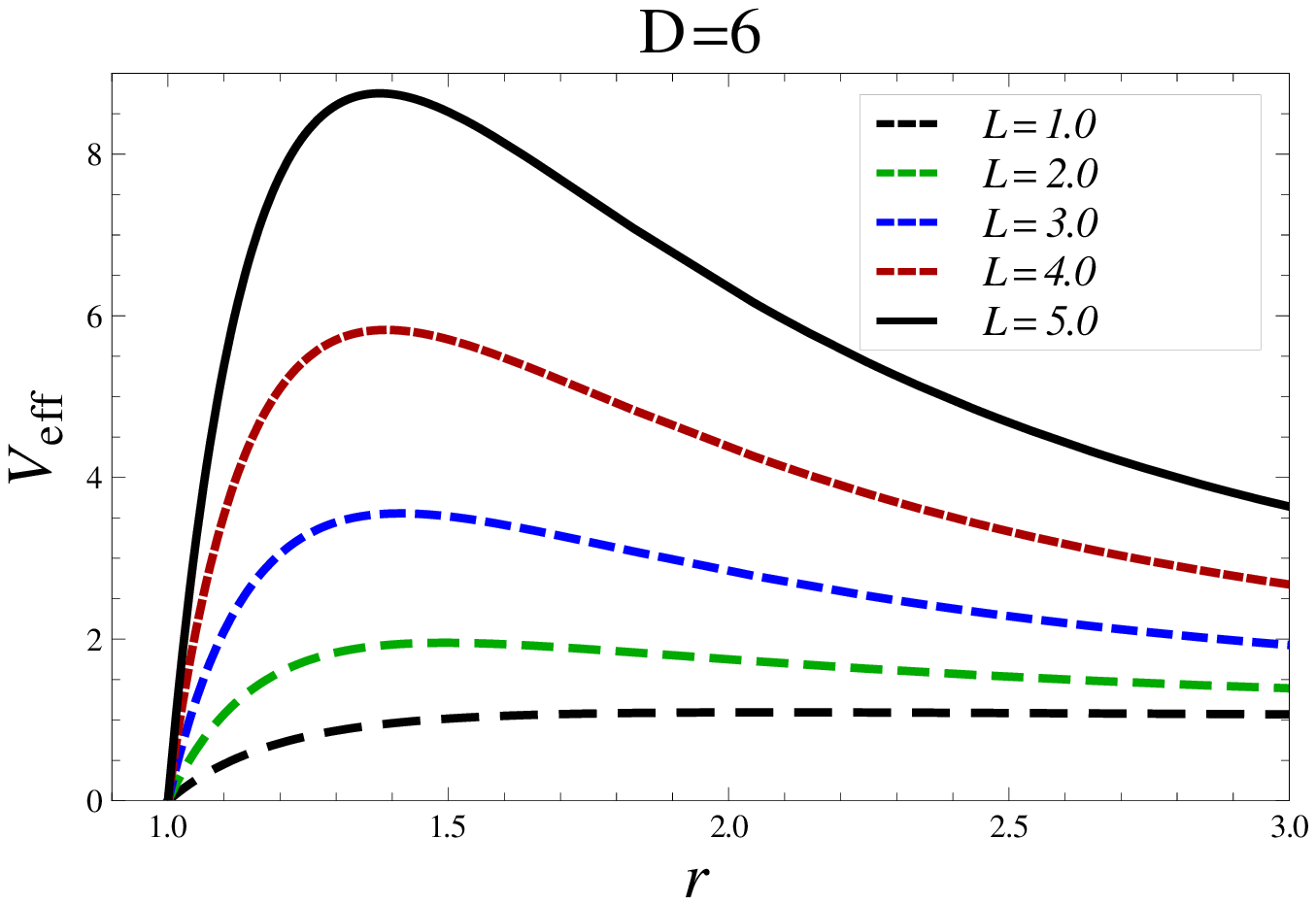} 
	\includegraphics[scale=0.6]{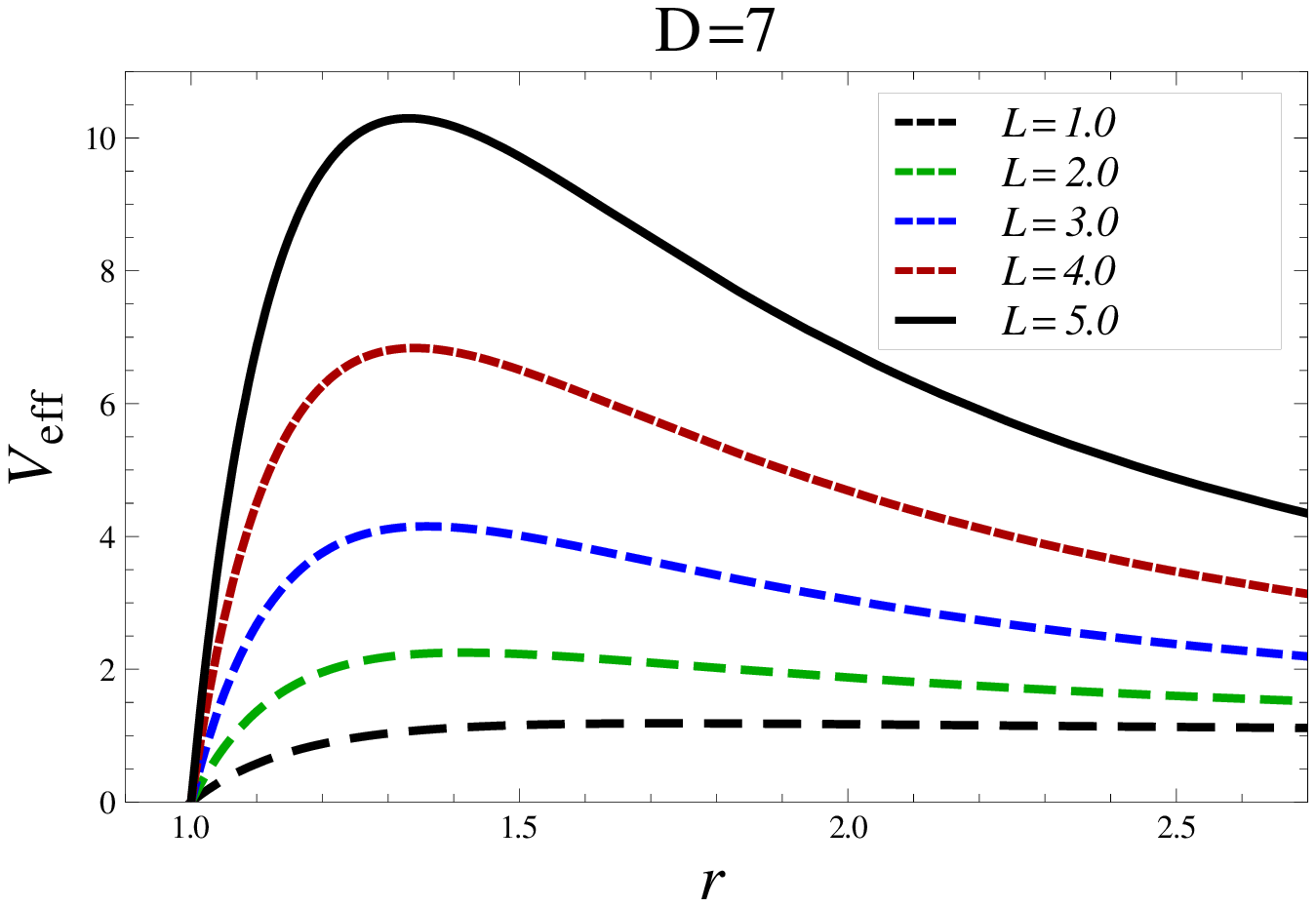} \\
	\includegraphics[scale=0.6]{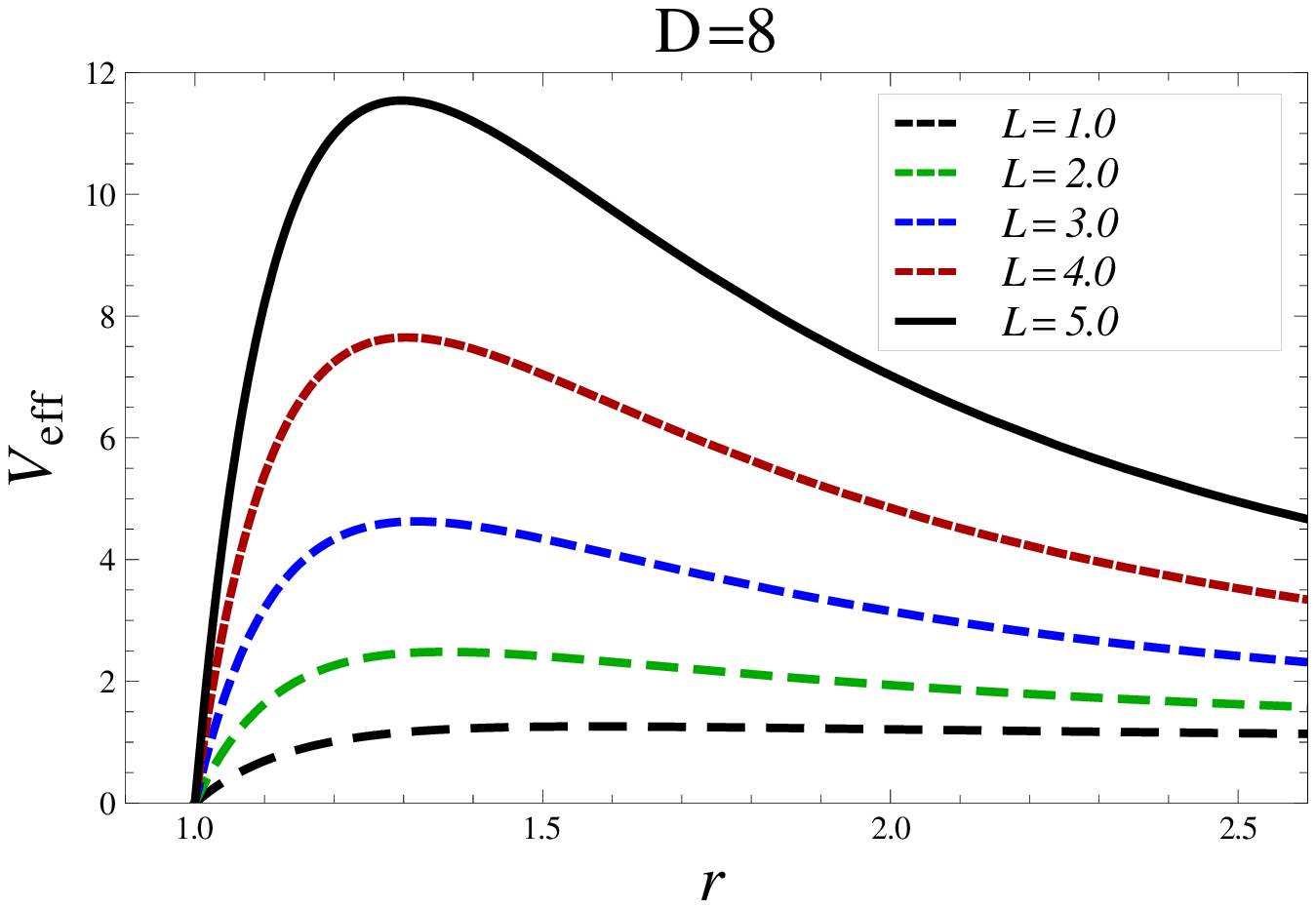} 
     \includegraphics[scale=0.6]{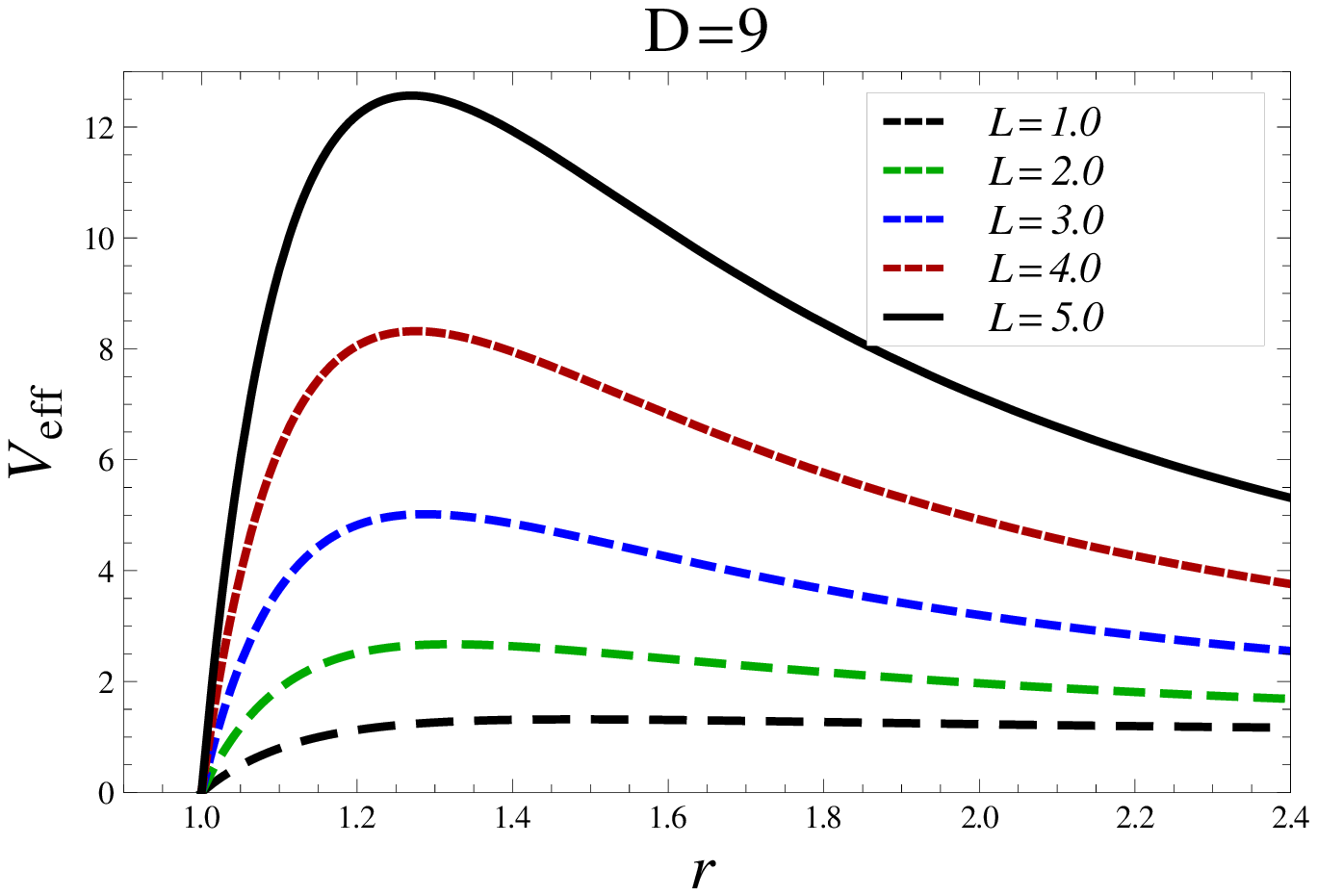} \\	
	\end{tabular}
    \caption{ Plot showing the dependency of effective potential on radial coordinate $r$ in different dimension $D$ and angular momentum $L$.}
\label{veff}
\end{figure*}
where $\mathcal{K}$ is the Carter constant \cite{carter}. By using Eqs.~(\ref{p1})-(\ref{p4}) in Eqs.~(\ref{sr}) and (\ref{stheta}), we get the complete null geodesics equation for Schwarzschild-Tangherlini black hole as
\begin{eqnarray}
&&\dot{t}=\frac{E}{1-{\frac{\mu}{r^{D-3}}}}, \label{tp}\\	&&\dot{\phi}=\frac{L}{r^2\displaystyle\prod\limits_{i=1}^{D-3}\sin^2\theta_i}, \label{p}\\
&&r^2\dot{r}=\pm\sqrt{\mathcal{R}},\label{r}\\
&&r^2\sum_{i=1}^{D-3}\prod_{n=1}^{i-1}\sin{\theta_i}\dot{\theta_i}=\pm\sqrt{\Theta_i}, \label{th}
\end{eqnarray}
where "$+$" and "$-$" sign respectively gives a motion of photon in outgoing and ingoing radial direction  and  a dot denote the derivative with respect to affine parameter $\tau$. For the  null curves the  expressions of  $\mathcal{R}(r)$ and $\Theta_i(\theta_i)$  in Eqs.~(\ref{r}) and (\ref{th}) can takes the form  as
\begin{eqnarray}
&&\mathcal{R}(r)= E^2r^4-r^2{\left[1-{\frac{\mu}{r^{D-3}}}\right]}\left[\mathcal{K}+{L}^2\right],\\\label{R}
&&\Theta_i(\theta_i)=\mathcal{K}-\prod_{i=1}^{D-3}{L}^2\cot^2\theta_i.
\end{eqnarray}
The Eqs.~(\ref{tp})-(\ref{th}) governs the motion of photon in Schwarzschild-Tangherlini space-time. The characteristics of photon near the black hole can be defined by two impact parameters, which are functions of constants of motion $E$, $L$ and $\mathcal{K}$. For the general orbits around the black hole the impact parameters are $\xi={L}/E$, $\eta={\mathcal{K}}/E^2$.
It is important to discuss the effective potential for determining the boundary of the shadow of a black hole. We can find out the effective potential by rewriting the radial null geodesic equation  for Schwarzschild-Tangherlini black hole which is given by
\begin{equation}
\left(\frac{d{r}}{d\tau}\right)^2+V_{eff}(r)=0,
\end{equation}
where $V_{eff}$ is the effective potential for radial motion given by  
\begin{equation}
V_{eff}=\frac{1}{r^2}\left(1-{\frac{\mu}{r^{D-3}}}\right)(\mathcal{K}+{L}^2)-E^2.\label{vef}
\end{equation}
Thus it is straight forward to show that the effective potential for Schwarzschild-Tangherlini is maximum for critical radius $r_c$ \cite{Decanini:2011xw}
\begin{equation}
r_c={\frac{1}{2^{D-3}}} \left[ \frac{16 \pi M(D-1)}{\Omega_{D-2} (D-2)}  \right]^{D-3}. 
\end{equation}
For $D$=4, the H$D$ equation of effective potential (\ref{vef}) reduces to Schwarzschild spacetime. We have plotted the radial dependency of effective potential in  Fig.~\ref{veff}. As we know for Schwarzschild case $V_{eff}$ for photon has maximum at $r=3M$, which shows a unstable circular orbit and as we go $r\rightarrow{\infty}$, effective potential asymptotes to a constant value. One can see from Fig.~\ref{fig2}, as we go to H$D$ the maximum value of effective potential increases, which emphasizes that as we go to the H$D$ unstable circular orbits becomes smaller. The photon orbits are circular and unstable corresponding to the maximum value of effective potential. The unstable circular orbit determine the boundary of apparent shape of the black hole and can be obtained by maximizing the effective potential, which demands
\begin{equation}
	V_{eff}=\frac{\partial V_{eff}}{\partial r}=0 \;\; \;\; \mbox{or}\;\;  R(r)=\frac{\partial R(r)}{\partial r}=0,\label{vr} 
\end{equation}
we obtain impact parameters $\eta$ and $\xi$ related to dimension $D$ via,
\begin{equation}
\eta +\xi ^2=\frac{D-1}{D-3} \left[ \frac{16{\pi}M(D-1)}{2(D-2)\Omega_{D-2}} \right]^{\frac{2}{D-3}}.\label{eta}
\end{equation}
The contour of the Eq.~(\ref{eta}) can describe the apparent shape of  Schwarzschild-Tangherlini black hole. For the Schwarzschild black hole the effective potential has maxima at $r_c=3M$.
\begin{table}
\centering
\caption{\label{tab:mbh} Variation of mass parameter $\mu$ and $\alpha^2$+$\beta^2$ with dimension $D$.}
\begin{tabular}{p{2.5cm}p{2.5cm}p{2cm}}
\hline
\hline
   $D$&          $\mu$    &    ${\alpha}^2+{\beta}^2$\\
\hline
    4 &                   $2M$             &               27$M^2$  \\
    5 &      \large{$\frac{8M}{3\pi}$}     &       \large{$\frac{32M}{3\pi}$}  \\
    6 &      \large{$\frac{3M}{2\pi}$}     &       \large{$\left(\frac{7M}{20\pi}\right)^\frac{2}{3}$} \\
    7 &      \large{$\frac{16M}{5\pi^2}$}  &       \large{$\left(\frac{108M}{5\pi^2}\right)^\frac{1}{2}$}  \\
    8 &      \large{$\frac{5M}{2\pi^2}$}   &       \large{$\left(\frac{10M}{\pi^2}\right)^\frac{2}{5}$}  \\
    9 &      \large{$\frac{48M}{7\pi^3}$}  &       \large{$\left(\frac{1446M}{\pi^3}\right)^\frac{1}{3}$}  \\
\hline
\hline
\end{tabular}
\end{table}
\begin{figure*}
    \begin{tabular}{c c c}
	\includegraphics[scale=0.51]{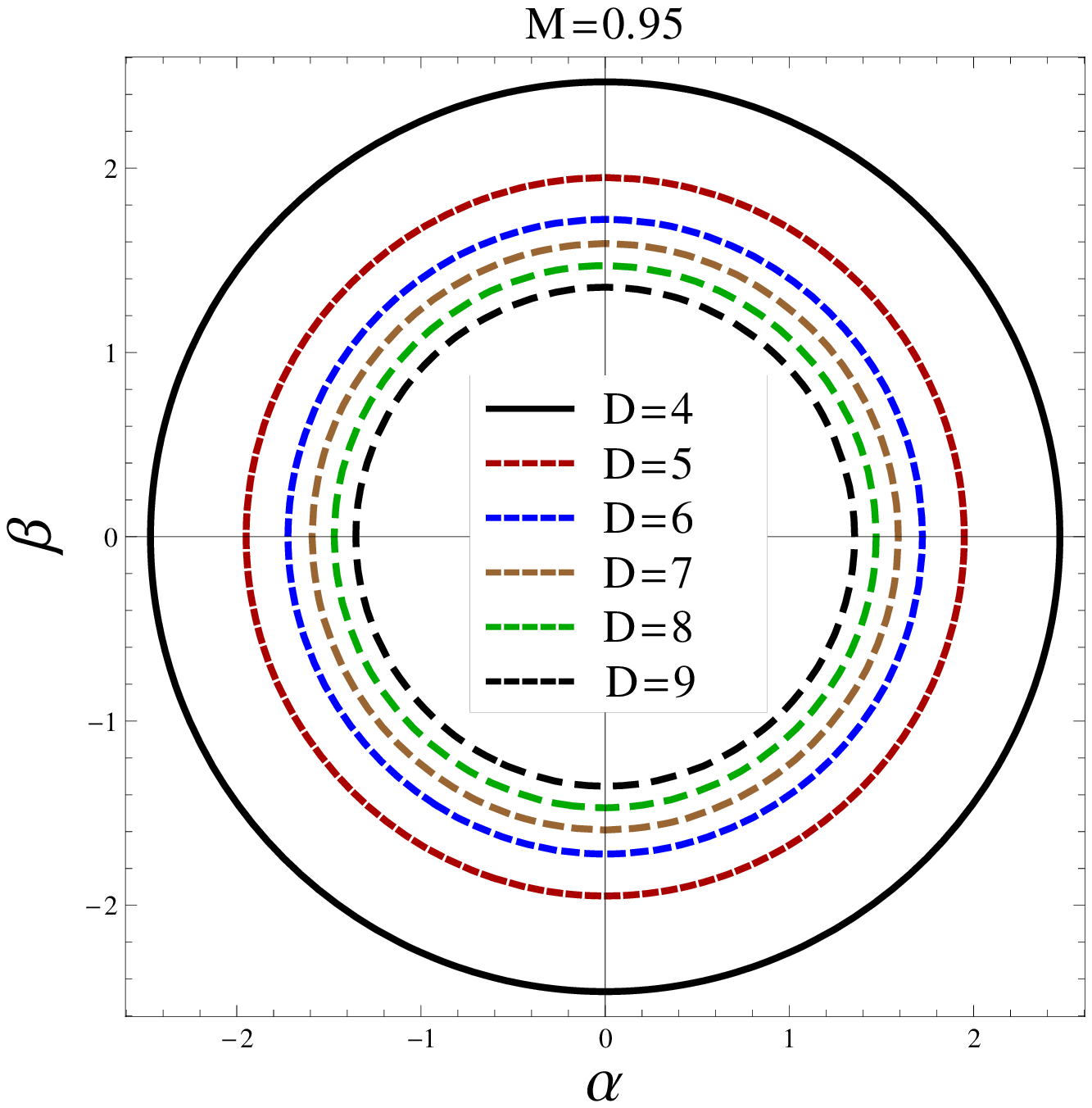} 
	\includegraphics[scale=0.51]{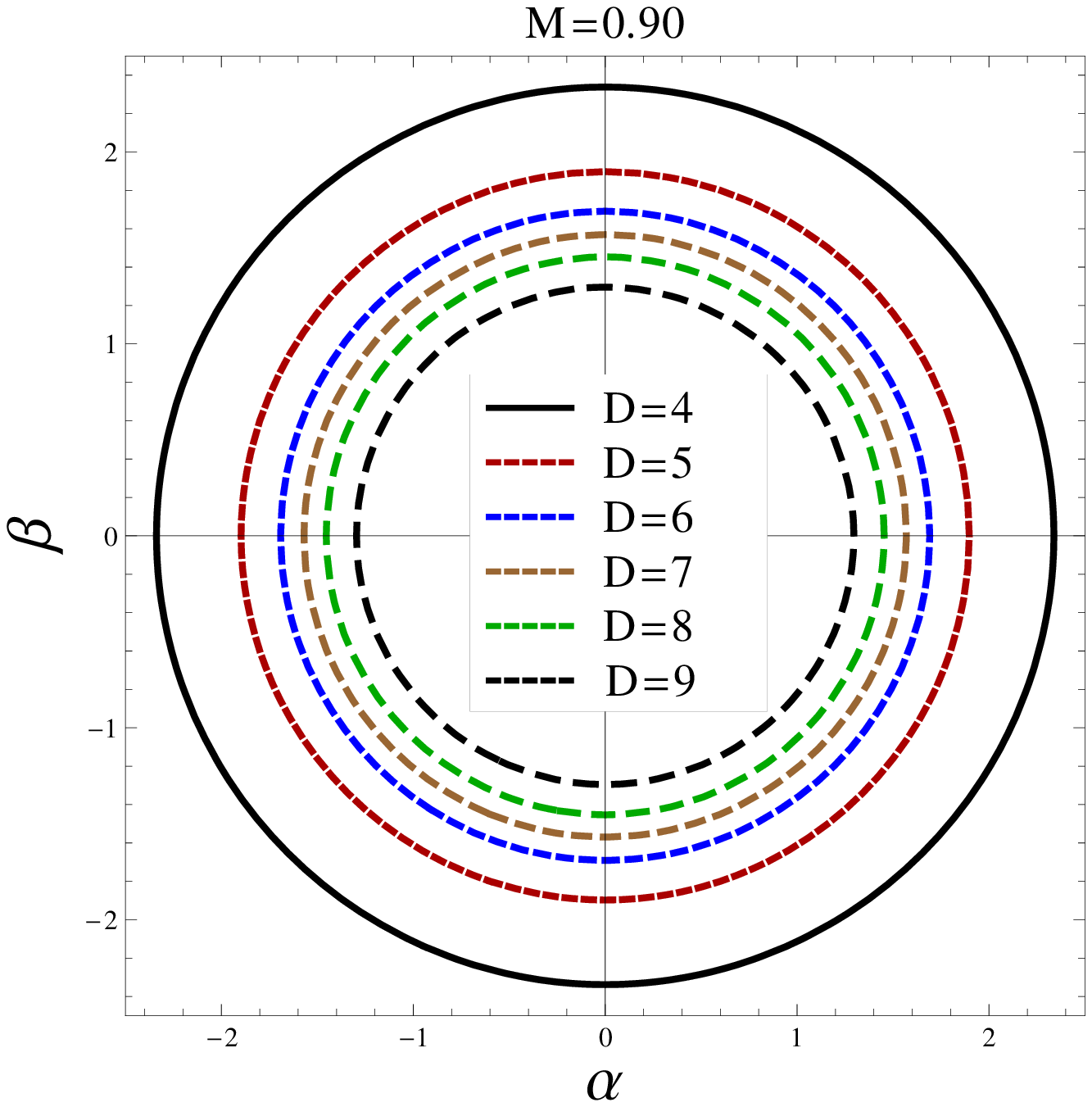} \\
     \includegraphics[scale=0.51]{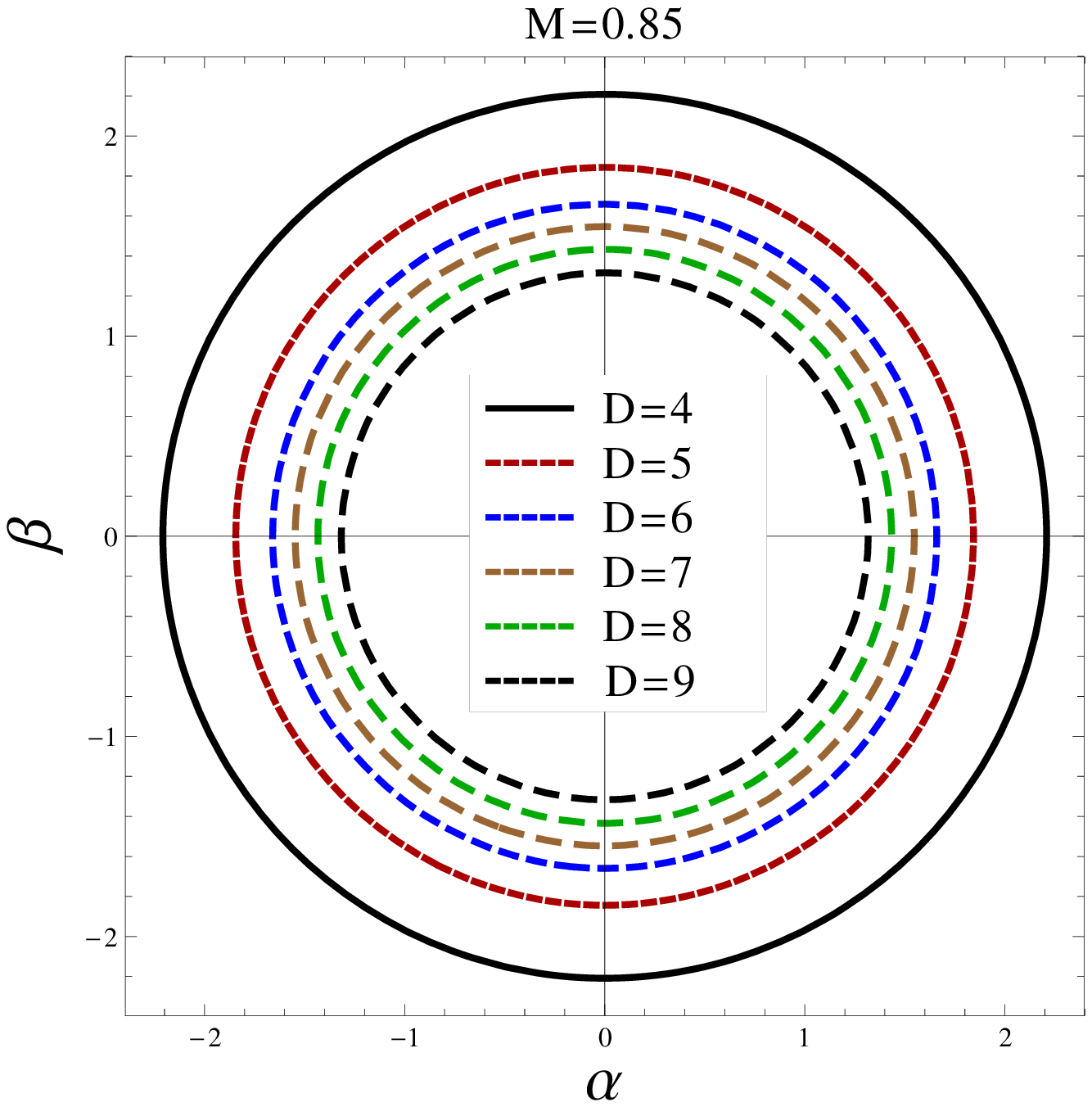}
	\includegraphics[scale=0.51]{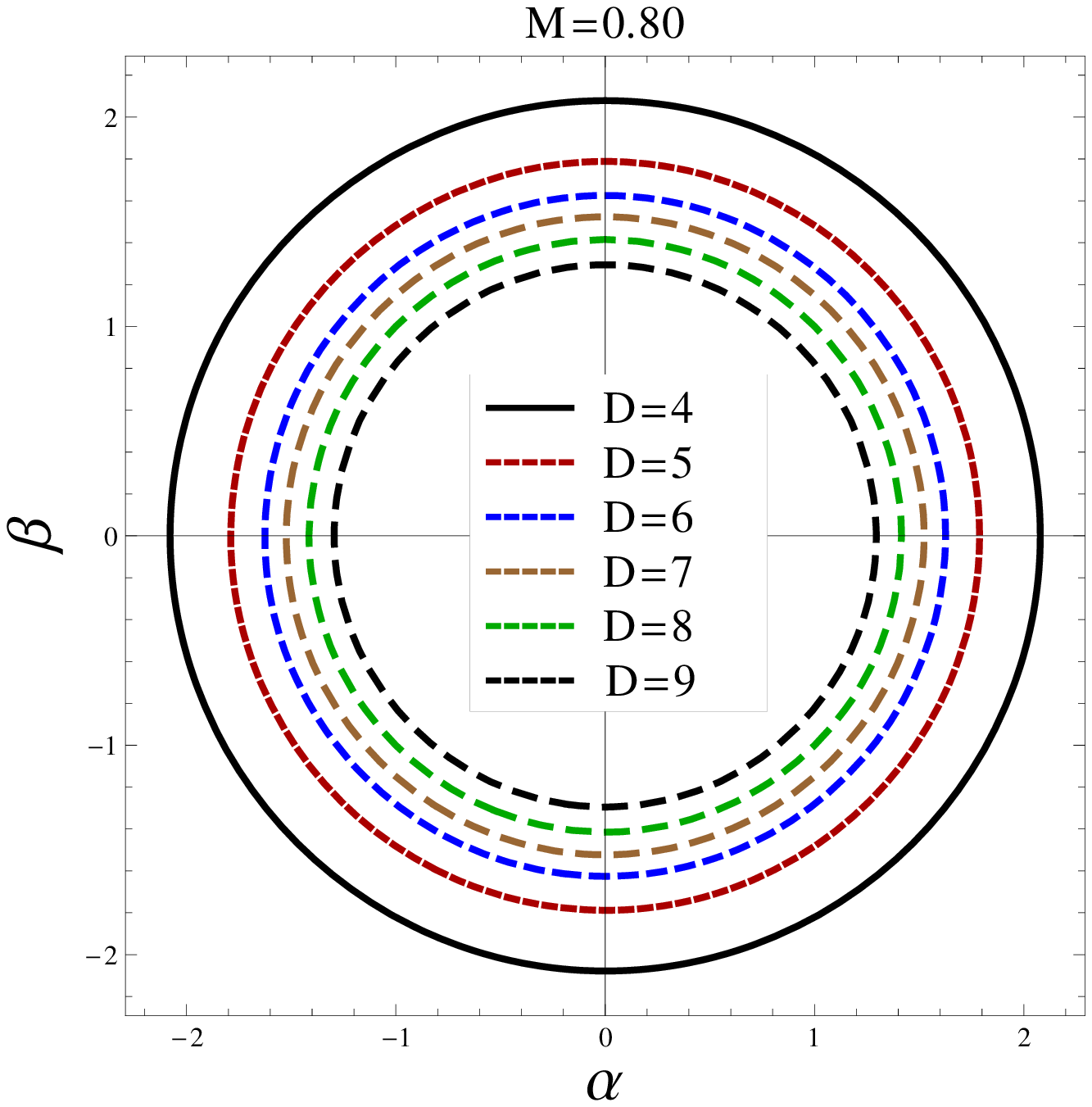}\\
	\includegraphics[scale=0.51]{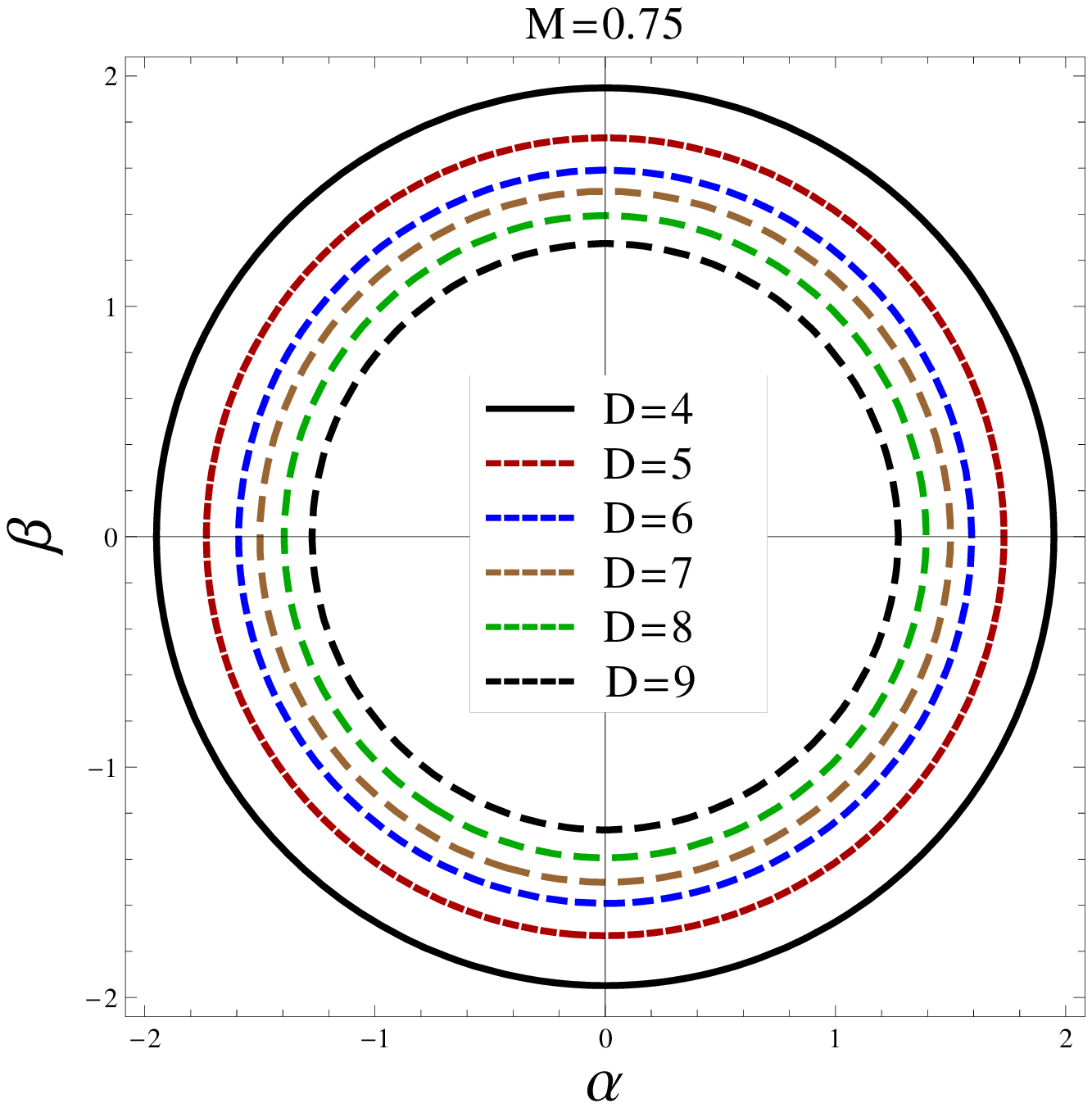} 
     \includegraphics[scale=0.51]{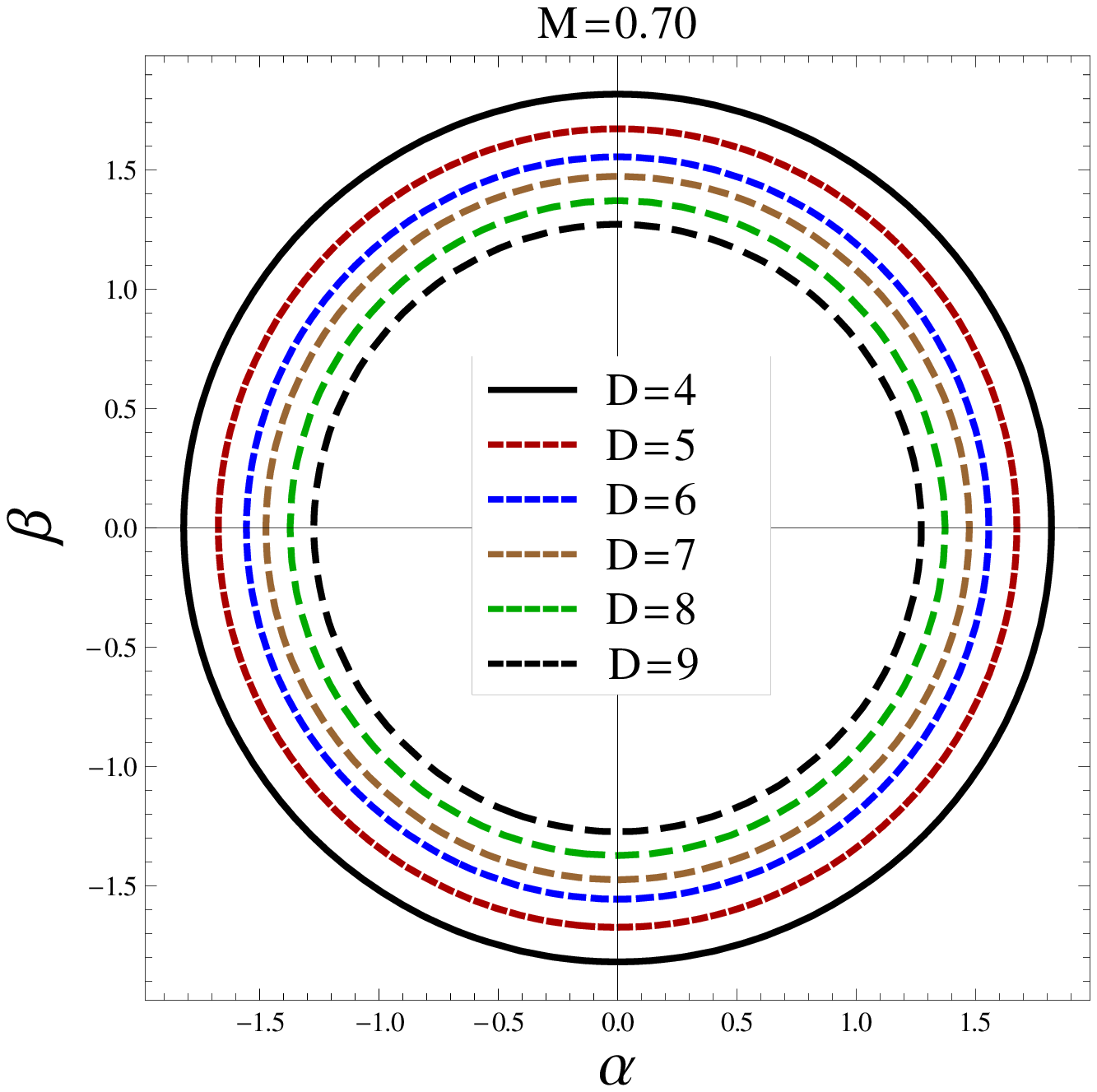} 
	\end{tabular}
         \caption{\label{fig2}    Shadow cast by Schwarzschild-Tanglerlini black holes at $\theta_i=\pi/2$, in different dimension $D$ and mass parameter $M$.}
\end{figure*}
\section{Shadow of The Black hole \label{sect4}}
The shadow of black hole and its photon orbit can be determined by geometrical optics. The apparent shape of a black hole is defined by the boundary of the shadow. From Eq.~(\ref{eta}) the size of  Schwarzschild-Tanglerlini black holes depends upon mass and dimension of space-time.  
To visualize the shadow of the black hole, it is appropriate to use the celestial coordinates $\alpha$ and $\beta$ \cite{cs}. For Schwarzschild-Tanglerlini black hole the celestial coordinate modified as \cite{Chandrasekhar:1992}
\begin{eqnarray}
&&\alpha=\lim_{r_0\rightarrow\infty}\left(\frac{r_0P^{(\phi)}}{P^{(t)}}\right),\label{al}\\
&&\beta_i=\lim_{r_0\rightarrow\infty}\left(\frac{r_0P^{(\theta_i)}}{P^{(t)}}\right) , \quad \:\:\:  i=1, . . . ,D-3,\label{be}
\end{eqnarray}
where $[P^{(t)}, P^{(\phi)}, P^{(\theta_i)}]$ are the vi-tetrad component of momentum. 
By using Eqs.~(\ref{p1})-(\ref{p2}) and geodesics equations of motion (\ref{tp})-(\ref{th}), one can obtain the expressions of celestial co-ordinates $\alpha$ and $\beta_i$, given by 
 \begin{eqnarray}
&&\alpha=\frac{-\xi}{\displaystyle\prod_{i=1}^{D-3}\sin\theta_i},\label{a}\\
&&\beta_i=\pm\sqrt{\eta-\xi^2\prod_{i=1}^{D-3}\cot^2\theta_i}.\label{b}
\end{eqnarray}
\begin{figure*}
    \begin{tabular}{c c c c}
	\includegraphics[scale=0.51]{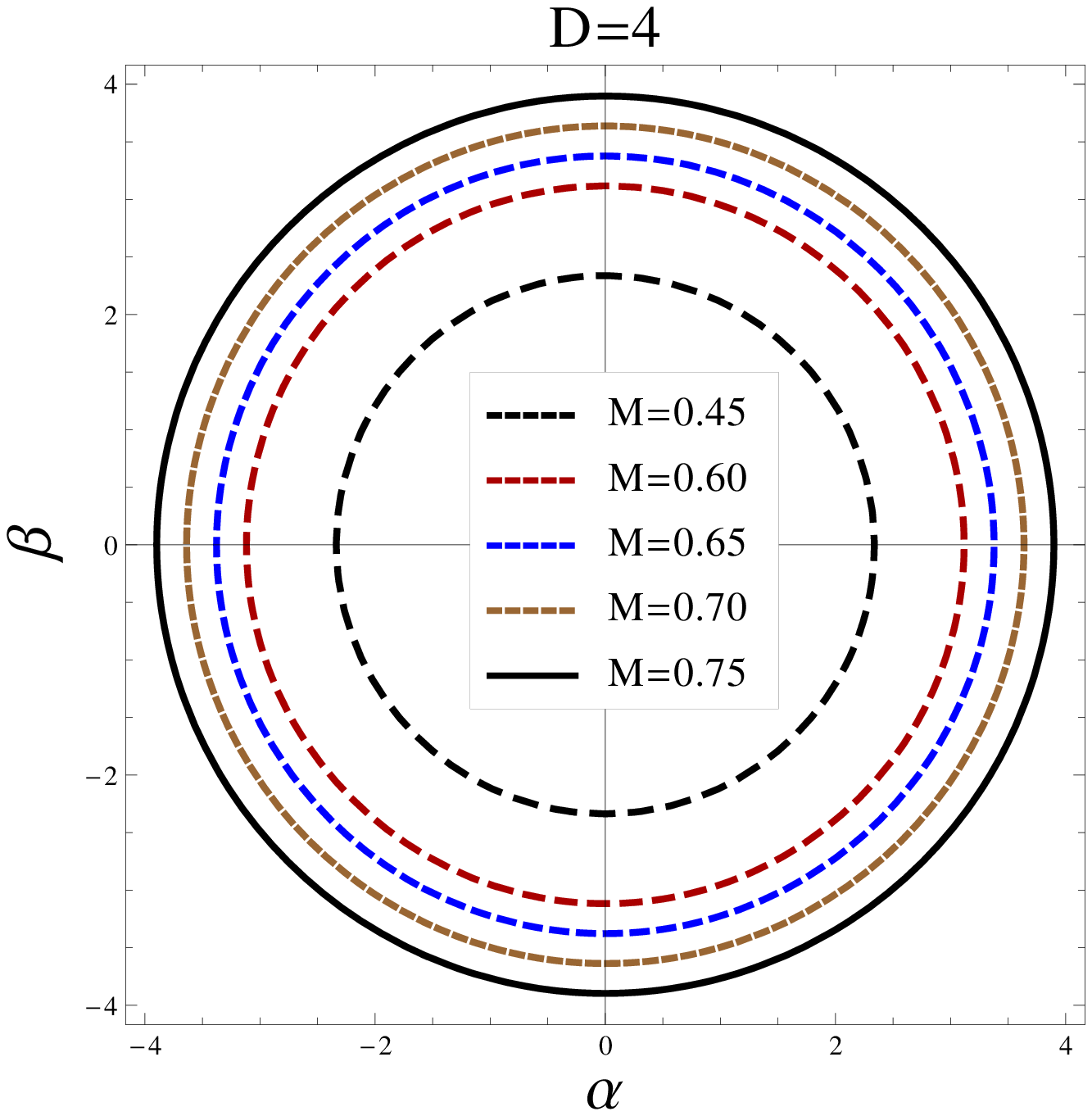} 
	\includegraphics[scale=0.51]{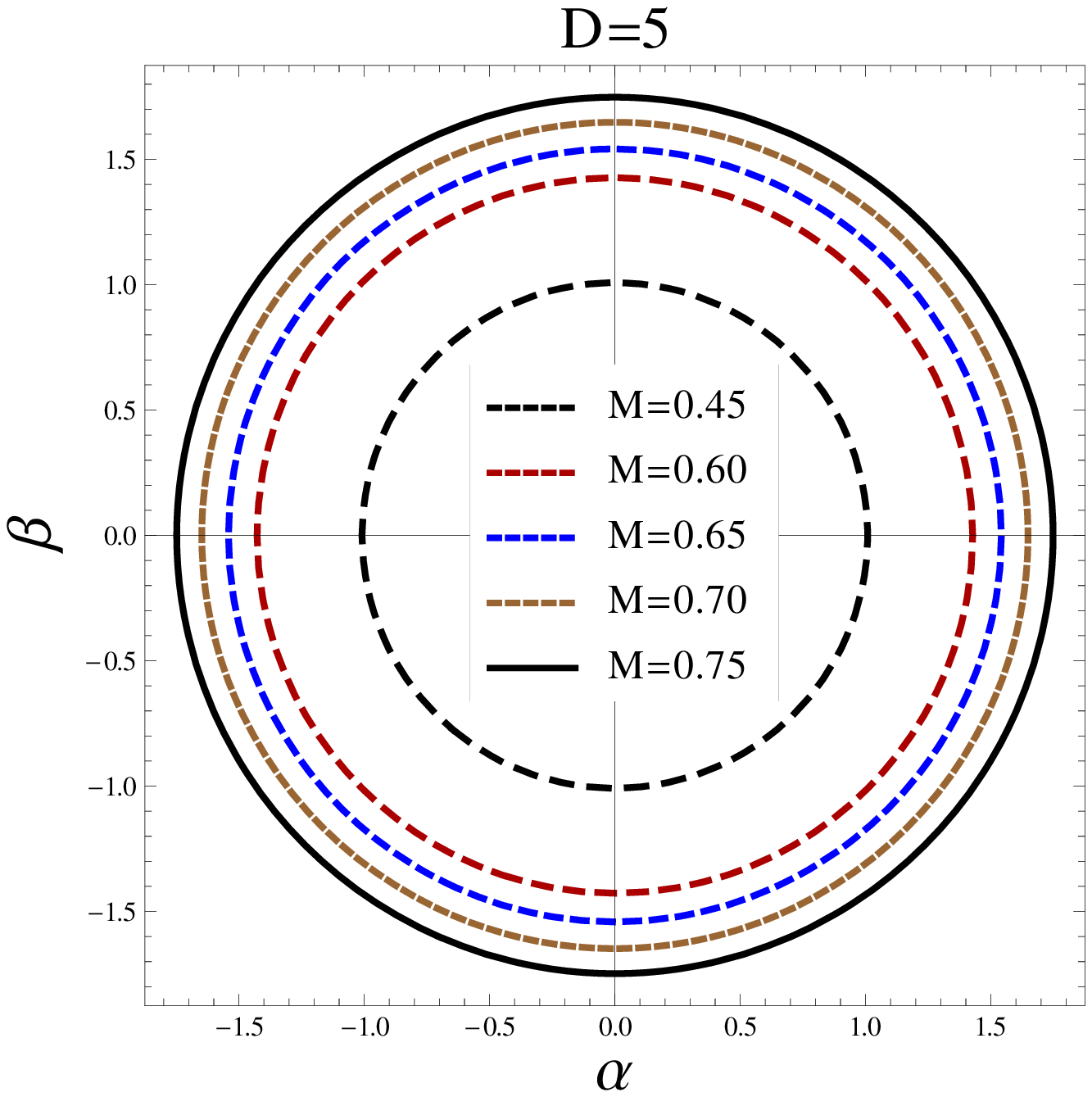} \\
     \includegraphics[scale=0.51]{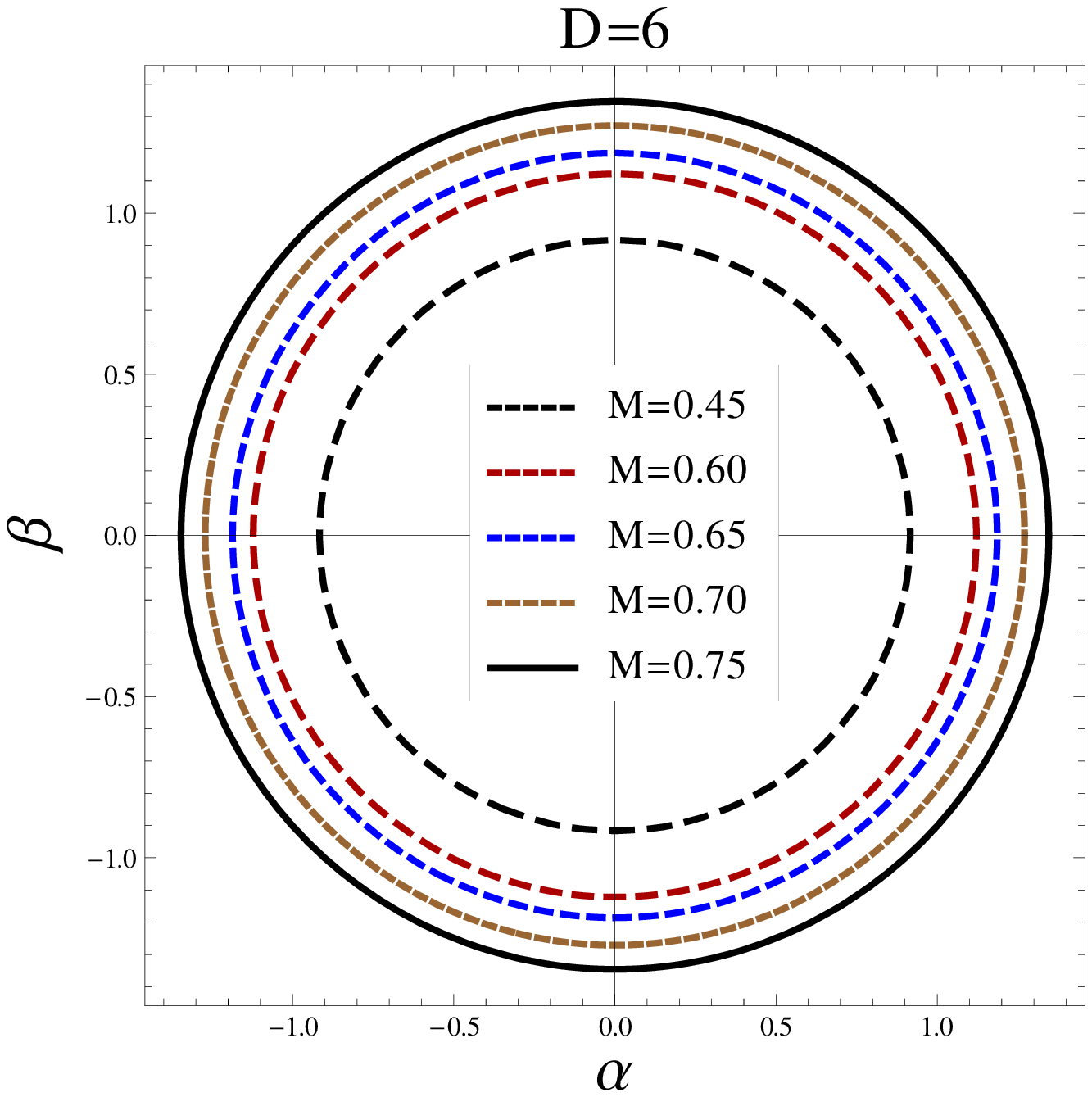}
	\includegraphics[scale=0.51]{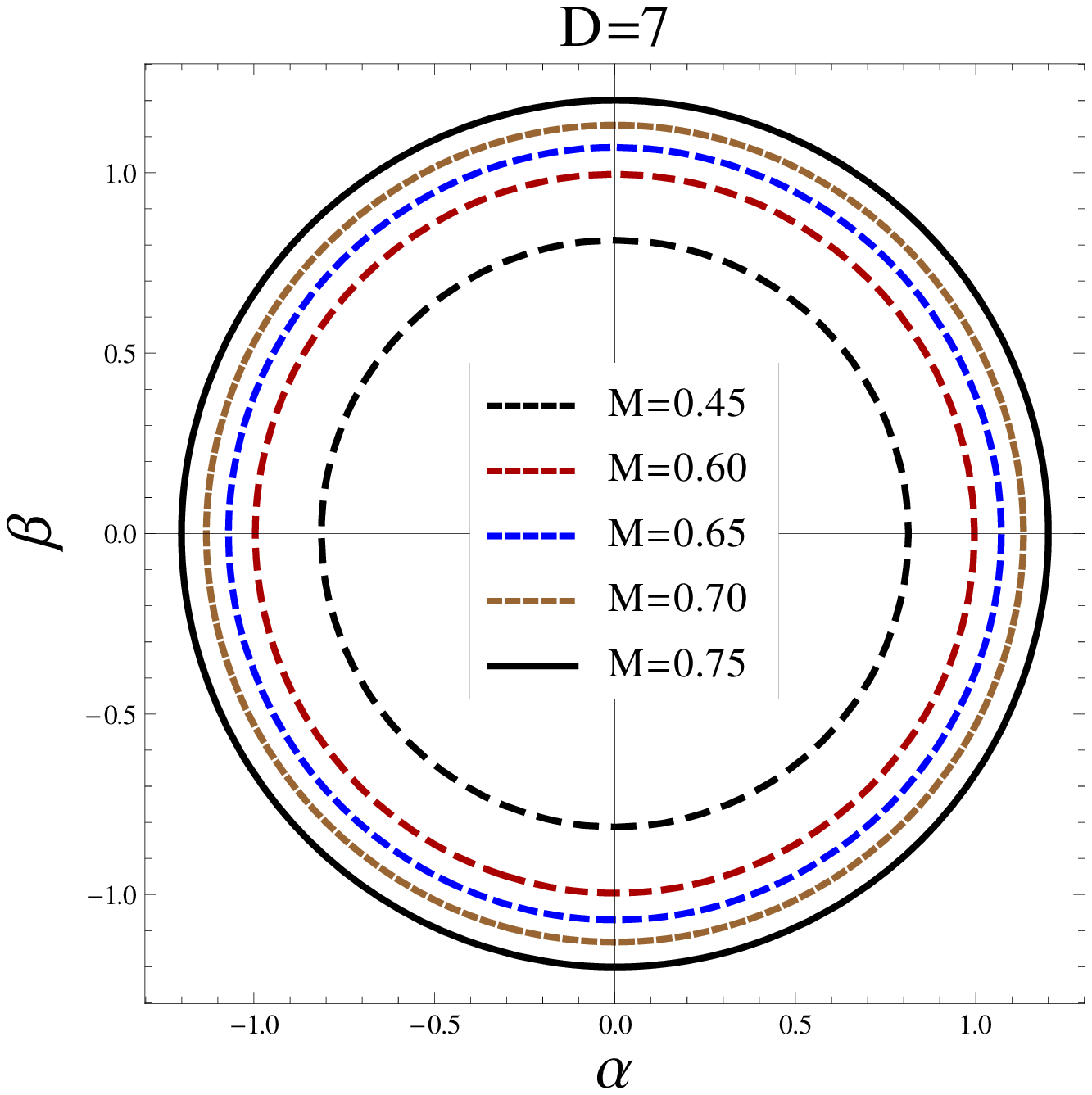} \\
	\includegraphics[scale=0.51]{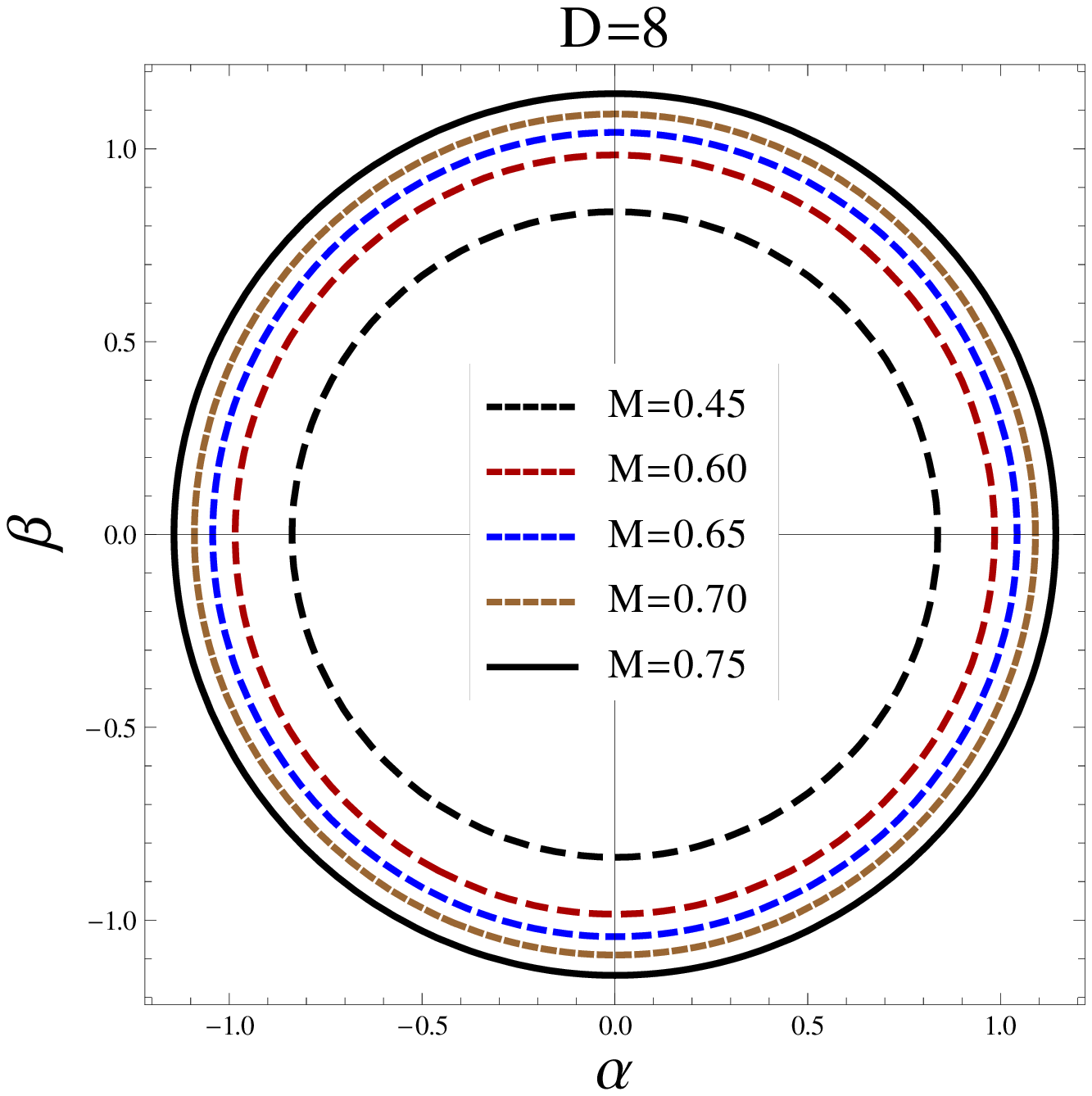} 
     \includegraphics[scale=0.51]{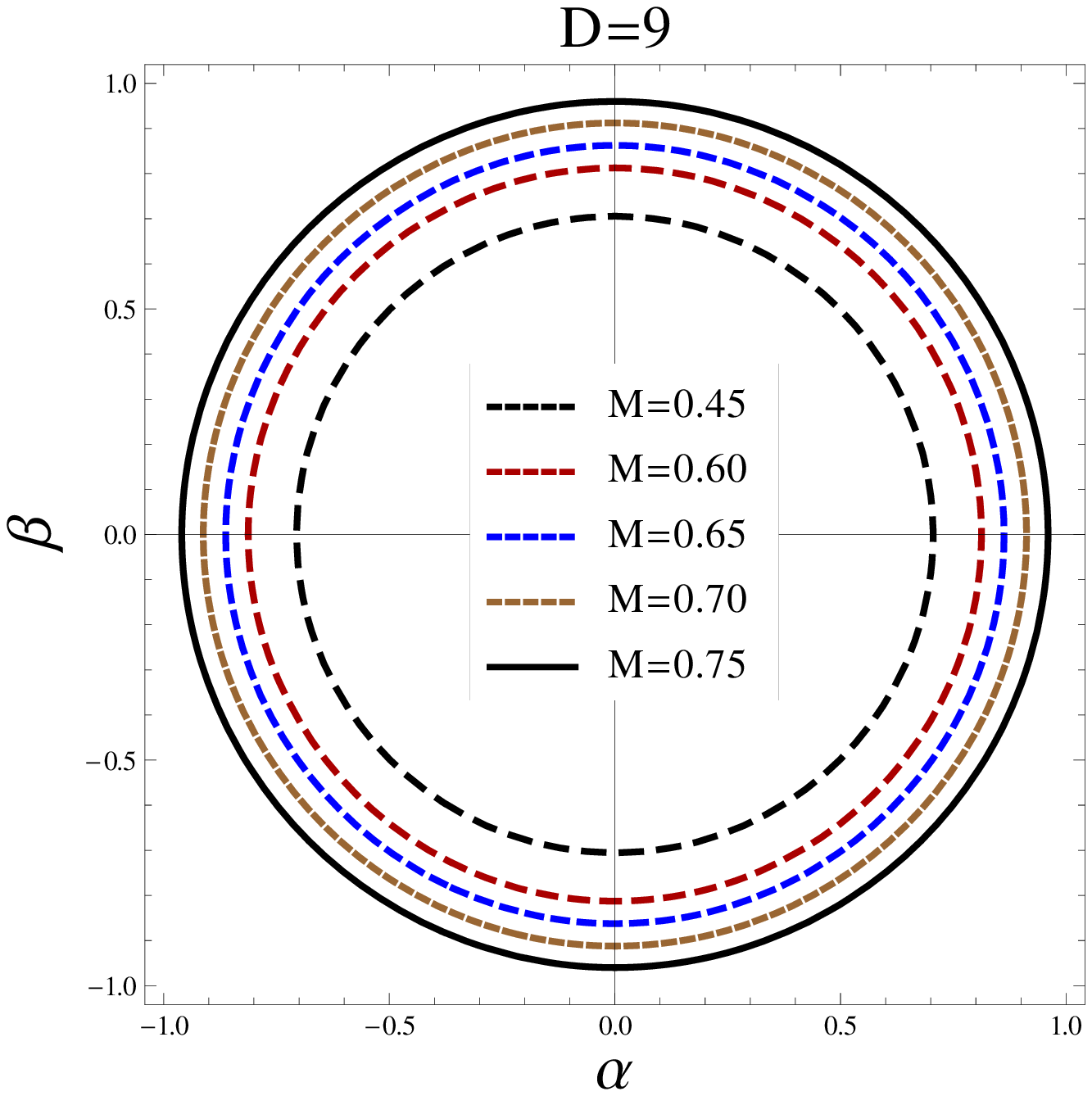} 	
	\end{tabular}
    \caption{\label{fig3}   Shadow cast by Schwarzschild-Tanglerlini black holes at $\theta_i=\pi/2$ for different  mass parameter $M$ and dimension $D$.}
\end{figure*}
The Eqs.~(\ref{a})-(\ref{b}) relates the celestial coordinate $\alpha$, $\beta_i$ to constant of motion $\eta$ and $\xi$. 
\begin{figure}
\begin{tabular}{ccc}
\includegraphics[width=0.65\linewidth]{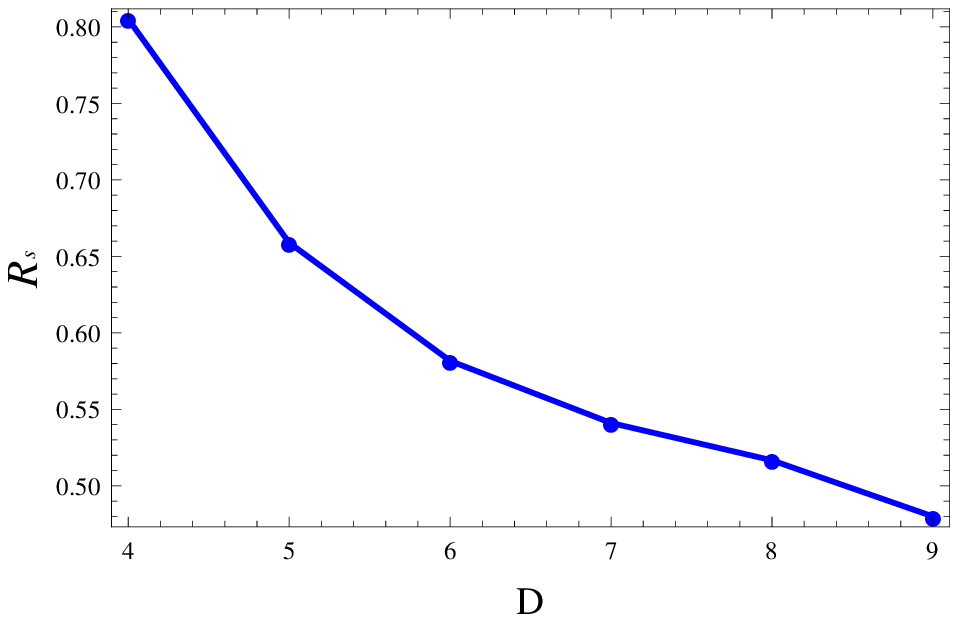}
\end{tabular}
 \caption{\label{fig4}  Plot showing the behavior of the radius of black hole shadow $R_s$ with dimension $D$.
}
 \end{figure}
If we take equatorial plane $(\theta_i =\pi/2)$, the celestial coordinate reduces to
\begin{equation}\label{alpha-beta1}
\alpha=-\xi, \;\;\;  \beta=\pm\sqrt{\eta}, 
\end{equation}
the Eq.~(\ref{alpha-beta1}) must follow the condition
\begin{eqnarray}
\eta +\xi ^2 = \alpha^2+\beta^2 = \frac{D-1}{D-3} \left[ \frac{16{\pi}M(D-1)}{2(D-2)\Omega_{D-2}} \right]^{\frac{2}{D-3}}. \label{n}
\end{eqnarray}
\begin{figure*}
    \begin{tabular}{c c c c}
\includegraphics[scale=0.54]{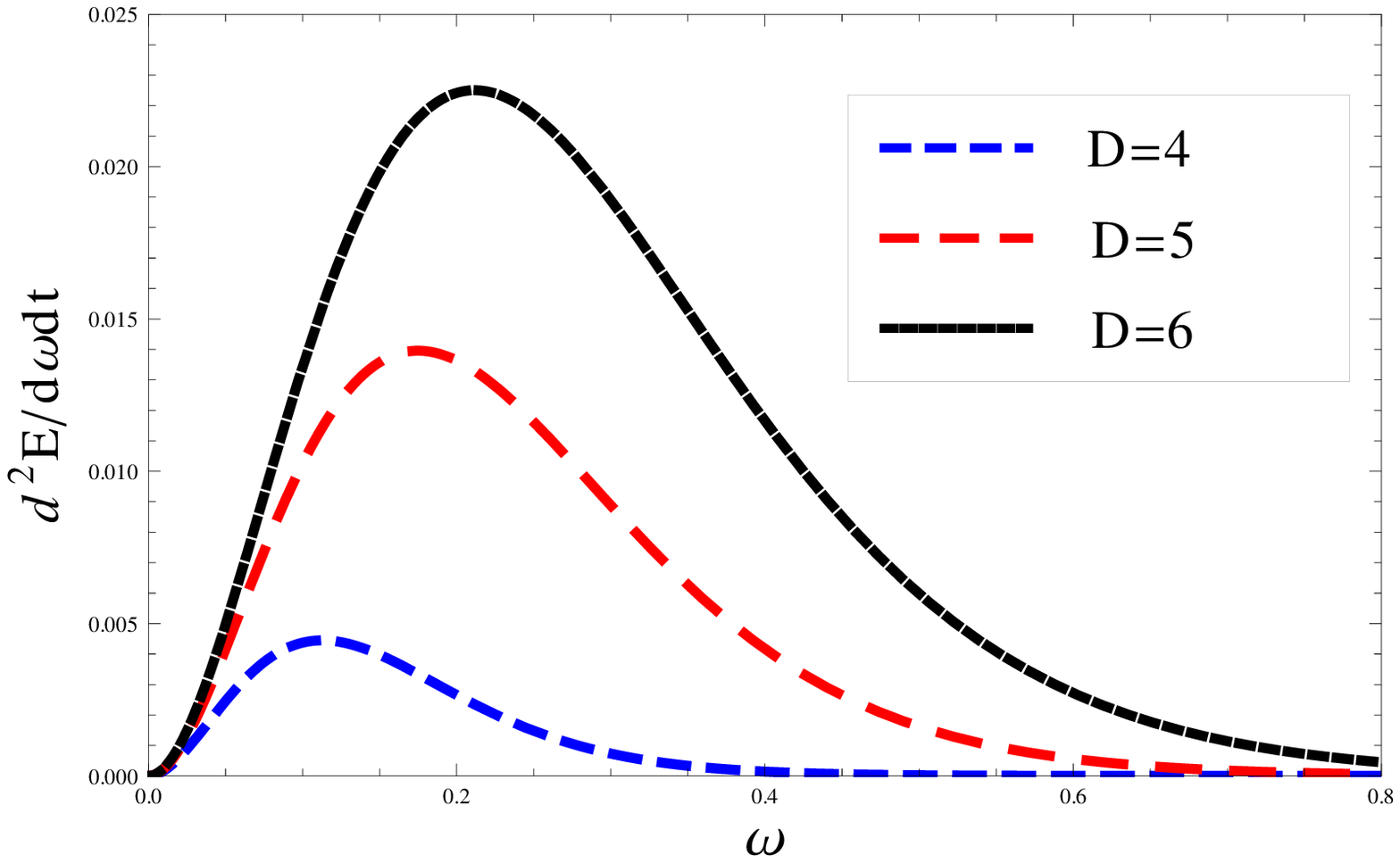}
\includegraphics[scale=0.54]{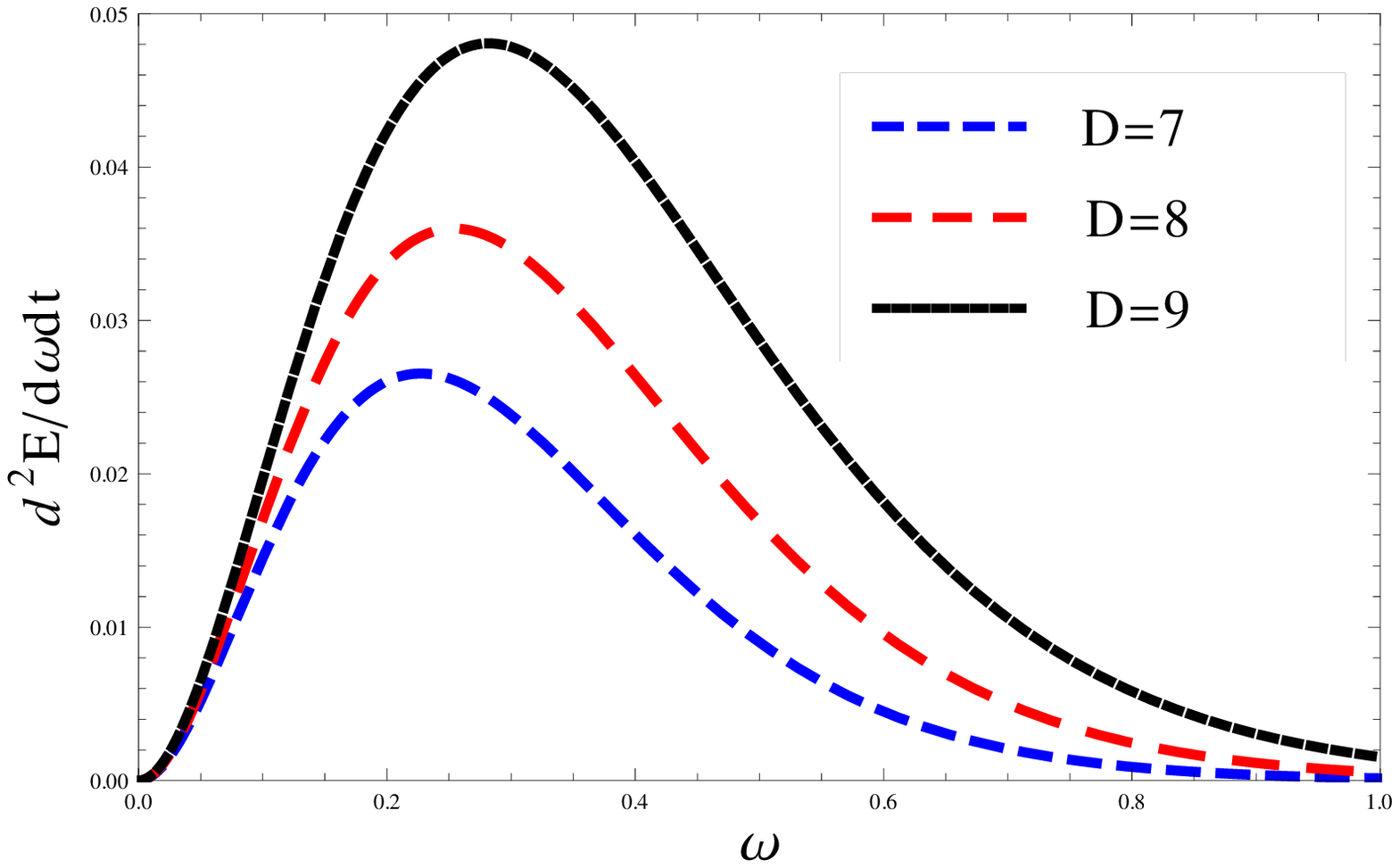}
\end{tabular}
 \caption{\label{fig6}  Plot showing the variation of  
 energy emission rate   for frequency $\omega$ in different dimension $D$.}
\end{figure*}
The Eq.~(\ref{n}) governs the complete orbit of photon around black hole which cast shadow and appears as circle. Now we take the contour plot of Eq.~(\ref{n}) which shows the shadow of Schwarzschild-Tangherlini black hole, clearly shown in Fig.~\ref{fig2} and \ref{fig3}. When we go to the H$D$ the effective size of shadow decreases, which physically represents a photon sphere decreases with the H$D$ of spacetime. For $D=4$, the Eq.~(\ref{n}) reduces to the most general case of a Schwarzschild black hole, reads
\begin{equation}
\eta +\xi ^2 = \alpha^2+\beta^2 = 27M^2,\label{t}
\end{equation}
and the contour plot of Eq.~(\ref{t})  represent shadow of Schwarzschild black hole which has been shown in first plot of Fig.~\ref{fig3}.

\subsection{Energy emission rate}
In this section, we study the energy emission rate from   Schwarzschild-Tangherlini 
black hole.
 The energy emission rate can be calculated via \cite{Wei:2013kza}
\begin{equation}
\frac{d^2E(\omega)}{d\omega dt}= \frac{2 \pi^2 \sigma_{lim}}{\exp{(\omega/T_{BH})}-1}\omega^3,
\end{equation}
where $T_{BH}$ is Hawking temperature for  Schwarzschild-Tangherlini black holes  which is shown in Eq.~(\ref{T}) and $\sigma_{lim}$ is the limiting constant value and in H$D$ it can be expressed as \cite{Decanini:2011xw, Decanini:2011xi}
\begin{equation}
\sigma_{lim}\approx  \frac{\pi^{\frac{(D-2)}{2}} R_s^{D-2}}{{\Gamma\left(\frac{D}{2}\right)}}, \nonumber
\end{equation}
for Schwarzschild case $(D=4)$ the value of limiting constant reduced to   
\begin{equation}
\sigma_{lim}\approx  \pi R_s^2, \nonumber
\end{equation}
where $R_s$ is radius of shadow. As we have shown in Fig.~\ref{fig2} and \ref{fig3}, the shadow of black hole is circle  and the radius of shadow $R_s$, which can also be a observable as \cite{Hioki:2009na}
 \begin{equation}
 R_s=\frac{(\alpha_t-\alpha_r)^2+\beta_t^2}{2(\alpha_t-\alpha_r)},
 \end{equation}
where $(\alpha_t,\beta_t)$ and $(\alpha_r,0)$ are the top, bottom and right positions of co-ordinates from where's reference circle passes. The complete form of energy emission of black hole in terms of dimensions $D$ can be reads 
 \begin{eqnarray}
\frac{d^2E(\omega)}{d\omega dt}= \frac{2 \pi^{\frac{D+2}{2}} R_s^{D-2}}{{(e^{ {\omega/T}} }-1){{\Gamma\left(\frac{D}{2}\right)}}}\omega^3. \nonumber
\end{eqnarray}
The variation of $d^2E(\omega)/{d\omega dt}$ vs $\omega$ can be seen from Fig.~\ref{fig6} for different dimension $D$.
\section{conclusion \label{sect6}}
The  black hole shadow in the near future may be realistic due to an observation of black hole Sgr$A^*$ in the center of our galaxy.  
It was Synge \cite{Synge:1966} and Luminet \cite{Luminet:1979} who first calculated the shadow of non-rotating Schwarzschild black hole, and demonstrated that circular light orbit exists on photon sphere at $r=3m$, which is the critical radius where the effective potential has a maximum.  We have generalized this and other results to study the shadow cast by Schwarzschild-Tangherlini black holes by studying the motion of a test particle and derive the complete null geodesics equations by applying Hamilton-Jacobi equation and Carter separable method. We also derive the expression of effective potential from the radial equation of motion and found that it has spacetime dimension dependence. It turns out that apparent shape of the Schwarzschild-Tangherlini black holes is also a function of spacetime dimension and that the size of black hole shadows increases with increase in the spacetime dimensions and so is energy emission rate. We also observe the deviation of the peak of effective potential towards the central object. 

The results presented here are the generalization of previous discussions, on the Schwarzschild black hole shadow, in more general setting, and the possibility of a further generalization of these results to H$D$ rotating Kerr black hole, Myers-Perry black hole and Lovelock black holes is an interesting problem for future research.
  
  \acknowledgements
 S.G.G. would like to thank SERB-DST Research Project Grant No. SB/S2/HEP-008/2014 and DST INDO-SA bilateral project DST/INT/South Africa/P-06/2016 and also to IUCAA, Pune for the hospitality while this work was being done. Special thanks to M. Amir for help in plots and also for fruitful discussions.

\end{document}